\documentclass[a4paper,11pt]{article}
\usepackage{jheppub} 
\usepackage{xcolor} 
\usepackage{amssymb}
\usepackage{dsfont}

\usepackage{ulem}
\usepackage{dutchcal}
\usepackage{bm}

\title{
An Algebraic Obstruction to Ising Criticality for Finite-BCH Entanglers in  Wavelet MERA}

\author[a,b]{Enrico Bertuzzo,}
\author[a,b]{Olindo Corradini }
\author[c]{and Claudia Frugiuele }

\emailAdd{enrico.bertuzzo@unimore.it, olindo.corradini@unimore.it, claudia.frugiuele@mi.infn.it}

\affiliation[a]{Dipartimento di Scienze Fisiche, Informatiche e Matematiche,\\
Università degli Studi di Modena e Reggio Emilia, Via Campi 213/A, I-41125 Modena, Italy}

\affiliation[b]{INFN, Sezione di Bologna, Via Irnerio 46, I-40126 Bologna, Italy}

\affiliation[c]{INFN, Sezione di Milano, Via Celoria 16, I-20133 Milano, Italy}

\abstract{
The computational treatment of non-Gaussian entanglers could pose a significant challenge when extending the Multi-Scale Entanglement Renormalization Ansatz (MERA) to interacting quantum field theories. A natural strategy is therefore to consider polynomial entanglers for which the Baker-Campbell-Hausdorff (BCH) expansion terminates at finite order.
In this work, we identify an algebraic obstruction that limits the universality classes accessible to this family of entanglers. Working within 
the wavelet MERA (wMERA) framework applied to the interacting 
$\phi^4$ theory in two dimensions, we show analytically that the effective potential 
generated by any finite-BCH polynomial entangler is necessarily of 
Landau form in the generic case, establishing mean-field universality 
for the full class; in non-generic, degenerate cases the resulting 
exponent departs from mean-field but still fails to reproduce the Ising 
value. As a numerical 
illustration, the critical exponent $\beta$ remains consistent with its 
mean-field value $\beta = 1/2$ across all ans\"{a}tze considered, with 
no drift toward the Ising value $\beta = 1/8$ as the nonlocality range 
or variational complexity increases.
Reproducing 
non-mean-field criticality therefore might require  
non-polynomial or infinite-BCH constructions.
}

\begin{document}
\maketitle

\section{Introduction}\label{sec:intro}

In recent years, there has been renewed interest in developing nonperturbative approaches to Quantum Field Theories (QFTs) beyond traditional lattice formulations ~\cite{BorelResum1,BorelResum2,phi4L1,HtruncationNLO,Vanhecke:2021noi,phi4DMRG,TNphi42}. A standard benchmark for such methods is the $(1+1)$-dimensional $\phi^4$ theory, which exhibits a nontrivial quantum phase transition in the Ising universality class while remaining sufficiently simple.
Tensor-network methods provide a promising route toward this goal. Matrix-Product States (MPS) \cite{mps1,mps2}  have been successfully adapted to interactig QFTs through various formulations \cite{TNphi41,TNphi42}. However, accurately capturing critical states with MPS generally requires a rapidly increasing bond dimension, reflecting the growth of long-range entanglement near criticality. This motivates the use of the Multiscale Entanglement Renormalization Ansatz (MERA) \cite{mera1}, whose hierarchical structure naturally incorporates scale-dependent entanglement and is closely connected to renormalization-group ideas.

An extension of MERA to QFTs, known as continuous MERA (cMERA), was 
introduced in Ref.~\cite{cMERA}, and an alternative wavelet-based 
formulation (wMERA) was subsequently developed in 
Ref.~\cite{wMERA} (see also \cite{waveletMERA1,quantumWavelet}), exploiting the 
multiscale structure and spatial localization of wavelets to encode 
entanglement efficiently \footnote{ More
recently, a related wavelet-based construction outside the MERA
framework has been
proposed, the wavelet matrix product states (wMPS),~\cite{tilloyw}, demonstrating the broader
utility of Daubechies wavelets as a discretization tool for
continuum field theories.}. While both frameworks have been successfully applied to free theories, extending them to interacting theories remains a significant challenge.
Existing attempts, developed exclusively within cMERA, focused mostly on implementing  non Gaussian entanglers to study
$(1+1)$-dimensional $\phi^4$ theory~\cite{IcMERA1,InteractingcMERA3,
InteractingcMERA4}.
Non-Gaussian entanglers with a finite Baker--Campbell--Hausdorff (BCH) expansion provide a natural starting point for interacting theories, as noted in Refs.~\cite{IcMERA1,InteractingcMERA3}. Their finite BCH structure makes them fully tractable within standard operatorial methods, providing a controlled setting to test whether the non-Gaussian correlations generated by this class of entanglers are already sufficient to reproduce the correct critical behavior.

In this work, we apply the wMERA construction of Ref.~\cite{wMERA}, based on D-6 Daubechies wavelets, to the interacting $\phi^4$ theory in $d=2$, providing the first  wMERA study of an interacting QFT. We then analyze the critical behavior of finite-BCH polynomial entanglers and identify an algebraic obstruction that constrains the universality classes accessible to this family of ansätze. We show that the effective potential 
generated by any finite-BCH polynomial entangler is  of 
Landau form in 
the generic case, so that the corresponding variational states are 
confined to the mean-field universality class, regardless 
of the nonlocality range or variational complexity of the ansatz. This is despite the fact that enlarging the variational manifold through increasingly non-local polynomial entanglers generates genuine non-Gaussian connected correlations. 
Our analytical results show that this expectation fails for a structural reason: the apparent increase in expressivity does not translate into new critical behavior. Indeed, the polynomial finite-BCH structure enforces an analytic effective potential of Landau form, preventing the emergence of Ising criticality.
We further consider a class of non-generic loopholes to this
argument, in which the general obstruction could in principle be
evaded; a Newton--Puiseux analysis shows that even in this case the
only accessible non-mean-field exponent is $\beta = 3/2$, which is also incompatible with the Ising value $\beta =
1/8$.  Moreover, we find no numerical evidence that such non-generic
configurations are ever physically realized across the full range of
ans\"atze considered.

The remainder of this paper is organized as follows. 
Section~\ref{intro} introduces the wavelet discretization of 
$\phi^4$ and the wMERA framework. Section~\ref{Epol} classifies polynomial 
non-Gaussian entanglers according to their BCH expansion, and establishes the 
general setup for the finite-BCH class.
Section~\ref{sec:obstruction} contains the main result: we show 
that finite-BCH polynomial entanglers are confined to a small set of 
non-Ising critical exponents, generically of Landau (mean-field) form, 
with a distinct exponent in non-generic, degenerate cases, and illustrate 
this analytically and numerically. 
Section~\ref{conclu} presents our conclusions.

\section{\texorpdfstring{Wavelet MERA for $\phi^4$ in $d=2$}{}}
\label{intro}
In this section we introduce the theoretical framework underlying our analysis. We briefly review the discretization of scalar field theory using 
Daubechies-6 (D-6) wavelets, following Ref.~\cite{wavelets}, to which 
we refer for further details. In the wavelet representation, continuum 
fields are decomposed into scaling and wavelet degrees of freedom by 
projecting them onto an orthonormal multiresolution basis.

The orthonormal scaling functions \(s_n^{r}(x)\) have compact support on the interval $n/2^r < x < (n+5)/2^r$ and capture coarse-grained, long-wavelength modes retained at resolution \(r\), while the orthonormal wavelet functions \(w_n^{r}(x)\) describe fluctuations localized at shorter lengths. We use the conventions of \cite{wMERA}, in which the scaling functions are defined in terms of a generating function $s(x)$ according to
\begin{equation}
    s_n^r(x) = \sqrt{2^r m_\phi} \,s(2^r m_\phi x - n).
    \label{eq:generating}
\end{equation}
Given a scalar field operator $\Phi(x)$ and its conjugate momentum $\Pi(x)$, the corresponding discrete field operators and conjugate momentum operators are defined as
\begin{subequations}\label{DOFphi}
\begin{align}
\phi_n^{r} &= \int dx\, \Phi(x) s_n^{r}(x), &
\varphi_n^{r'} &= \int dx\, \Phi(x) w_n^{r'}(x),\\
\pi_n^{r} &= \int dx\, \Pi(x) s_n^{r}(x), &
\varpi_n^{r'} &= \int dx\, \Pi(x) w_n^{r'}(x).
\end{align}
\end{subequations}
These operators satisfy the standard canonical commutation relations
$
[\phi_n^{r}, \pi_m^{r'}] = i\delta^{rr'}\delta_{nm},$ and 
$ 
[\varphi_n^{r}, \varpi_m^{r'}] = i\delta^{rr'}\delta_{nm},$
reflecting the orthonormality of the scaling and wavelet bases. 
Wavelet coefficients replace field amplitudes as the natural variables of the theory, with fields at resolution $r$ decomposed as
\begin{equation}\label{eq:field_decomposition}
    \Phi^r(x) = \sum_n \phi_n^r\, s_n^r(x) = \sum_n \left[\phi_n^{r-1}\,s_n^{r-1}(x) + \varphi_n^{r-1}\, w_n^{r-1}(x)  \right]~,
\end{equation}
and similarly for the conjugate momentum operator. The conversion of scaling and wavelet functions at layer $r-1$ to pure scaling functions at layer $r$ is called a wavelet transform~\cite{wavelets,wMERA}.

It is useful to define creation and annihilation operators
\begin{subequations}\label{aadagger}
\begin{align}
\phi_n^{r} &= \frac{1}{\sqrt{2\Delta}}(\sigma_n^{r\,\dagger}+\sigma_n^{r}), &
\pi_n^{r} &= i\sqrt{\frac{\Delta}{2}}(\sigma_n^{r\,\dagger}-\sigma_n^{r}),\\
\varphi_n^{r} &= \frac{1}{\sqrt{2\Delta}}(\omega_n^{r\,\dagger}+\omega_n^{r}), &
\varpi_n^{r} &= i\sqrt{\frac{\Delta}{2}}(\omega_n^{r\,\dagger}-\omega_n^{r}),
\end{align}
\end{subequations}
which, by construction, satisfy
$
[\sigma_n^{r},\sigma_m^{r'\,\dagger}] = \delta^{rr'}\delta_{nm}, $ and  $
[\omega_n^{r},\omega_m^{r'\,\dagger}] = \delta^{rr'}\delta_{nm}.
$
The scale parameter $\Delta$ is, in principle, arbitrary. In what follows, we set $\Delta = m_\phi$, having verified that variationally optimizing it does not modify the results presented in later sections.

We now briefly summarize the wMERA framework, closely following Ref.~\cite{wMERA}. We define two characteristic scales: the IR scale ($\mathcal{r}$), and the UV scale $(\mathcal{R})$, which serves as the ultraviolet cutoff of the effective theory. The main idea is to construct an approximation for the vacuum state at the UV scale starting from a simple ansatz for the IR vacuum, taken to be a product state
\begin{equation}
\bigl|\mathbf{P}_{\!\sigma}\bigr\rangle_{\mathcal{r}} \equiv \prod_n \bigl|\mathbf{0}_{\sigma_n}\bigr\rangle_{\mathcal{r}},
\end{equation}
where $\bigl|\mathbf{0}_{\sigma_n}\bigr\rangle_{\mathcal{r}}$ is the vacuum annihilated by the scaling operator $\sigma_n^\mathcal{r}$. 
This is achieved by entangling neighboring sites of the product-state vacuum with a unitary entangler $U_\mathcal{r}$ and then adding layers at different resolutions. Each layer is obtained by enlarging the Hilbert space adding wavelets degrees of freedom through a fine-graining map $W_r$, entangling the scaling and wavelet degrees of freedom with a quasi-local unitary $V_r$, and applying the inverse wavelet transform $T_r^{-1}$ to return to a description in terms of scaling modes only. Repeating this procedure up to the UV scale $\mathcal{R}$ gives
\begin{equation}
\label{vuotowMERA}
\big|\mathbf{0_\sigma}\rangle_\mathcal{R}
=\Bigg[ \prod_{r=\mathcal{r}}^{\mathcal{R}} 
{T}_r^{-1} {V}_r {W}_r \Bigg]U_\mathcal{r}
\big|\mathbf{P}_{\!\sigma}\rangle_{\mathcal{r}}. \qquad
\end{equation}
In general, the variational parameters of the entangling unitaries are fixed by minimizing the vacuum expectation value of the Hamiltonian at each layer.





 
Computations are  conveniently carried out  in the ``Heisenberg'' picture, i.e.\ considering the 
transformed fields $\phi_n^r \to U^\dagger_r \,\phi_n^r\, U_r$ 
(with an analogous transformation holding for the conjugate momenta). 
Field operators are thus written in terms of creation and annihilation 
operators (cfr. Eq.~\eqref{aadagger}), which satisfy canonical 
commutation relations. This, together with the fact that expectation values are taken over product states, implies that Wick's theorem can be readily applied. This operatorial framework is 
naturally suited to non-Gaussian entanglers admitting a finite BCH 
expansion, for which the transformed fields remain polynomial and 
Wick's theorem applies directly; their critical properties are 
analyzed in detail in the following Sections.

Let us now consider the wavelet discretization of the Hamiltonian in $\phi^4$ in $d=2,$ that is
\begin{equation}
\label{H(x)}
H = \int dx\,\Bigg[\,\frac{1}{2}\Pi^2(x) + \frac{m_\phi^2}{2}\Phi^2(x) + \frac{1}{2} (\nabla\Phi(x))^2 + \lambda \Phi^4(x)  \Bigg],
\end{equation}
where it is useful to define the dimensionless coupling $g \equiv \frac{\lambda}{m_\phi^2}.$
Inserting the decomposition\,\eqref{eq:field_decomposition} (and the analog for the momentum operator $\Pi(x)$) at resolution $r$ in the Hamiltonian\,\eqref{H(x)} we obtain:
\begin{equation}
\label{Hrtot}
H^r_\phi = \frac{1}{2} \sum_i \Big[ (\pi_i^r)^2 + m_\phi^2 (\phi_i^r)^2 + (2^r m_\phi)^2 \sum_j \phi_i^r \mathds{K}_{ij}^{ss} \phi_j^r \Big] + \mathcal{V}_\phi^r[\phi^r]~,
\end{equation}
with 
\begin{align}
   \mathds{K}_{ij}^{ss} =\int dy\, \nabla s(y-i) \cdot \nabla s(y-j)~,
\end{align}
where $y$ is a dimensionless variable and $s(x)$ is a generating function---see Eq.~\eqref{eq:generating}. The $\mathbb{K}_{ij}^{ss}$ coefficients vanish for $|i-j|\ge5$ due to our choice of decomposing the field into D-6 Daubechies wavelets.
The quartic interaction is
\begin{equation}
\label{Ltensor4}
\mathcal{V}_\phi^r[\phi^r] = 2^r  g\,  m_\phi^3 \sum_{i_1,i_2,i_3,i_4} \Lambda_{i_1 i_2 i_3 i_4} \phi_{i_1}^r \phi_{i_2}^r \phi_{i_3}^r \phi_{i_4}^r, \quad
\Lambda_{i_1 i_2 i_3 i_4} = \int dy\, s(y-i_1)\dots s(y-i_4),
\end{equation}
with $\Lambda_{i_1 i_2 i_3 i_4} \neq 0$ only if $|i_a - i_b| \le 4$. Again, this finite-range is induced by the compact support of D-6 scaling functions.

Since we are interested in studying the theory near criticality, a fundamental quantity to consider is the vacuum expectation value (vev) of the field $\Phi$. This can be computed according to
\begin{equation}\label{veve}
    \langle \Phi^r \rangle = \sum_n \langle \phi_n^r \rangle\, s_n^r(x),
\end{equation}
from which we see that a translationally invariant vev (i.e. $x$-independent) can be obtained only if $\langle \phi_n^r \rangle$ does not depend on $n$, since in this case we can use the identity $\sum_n s_n^r(x) = \sqrt{2^r m_\phi}$\,\cite{wavelets,wMERA}. The vev we are considering is taken over the $|0_\sigma \rangle_r$ state, i.e.
\begin{equation}
    \langle \phi_n^r \rangle \equiv \langle 0_\sigma | \phi_n^r | 0_\sigma \rangle_r = \langle \mathbf{P}_{\!\sigma}| U_r^\dagger \phi_n^r U_r |\mathbf{P}_{\!\sigma} \rangle_{\mathcal{r}}.
\end{equation}

\section{Finite-BCH Polynomial Entanglers}
\label{Epol}


We consider non-Gaussian entanglers generated by monomial operators,

\begin{equation}
\label{Ui}
    U_{j} = e^{i K_j[\phi, \pi]} \,,
\end{equation}
where  $K_j$ takes the form
\begin{equation}\label{eq:Ks}
    K_1 =  \mathbf{T}\,\pi^A \phi^B + h.c. \,, \qquad
    K_2 = \mathbf{T}\,\pi^C \,, \qquad
    K_3 =  \mathbf{T}\,\phi^D \,,
\end{equation}
with $A+B,C,D > 2$. Here, $\mathbf{T}$ denotes a tensor (of order $A+B$, $C$, and $D$, respectively) constructed from variational parameters, which must satisfy appropriate symmetry conditions and, by construction, is invariant under linear shifts of the indices of $T$. For instance, a simple possibility with $A=1$ and $B=2$ is $K_1 = T_{ijk}\, \phi_i\,\phi_j\,\pi_k + h.c.$, in which case $T_{ijk}$ is symmetric in $i \leftrightarrow j$ and satisfies $T_{ijk} = T_{i-m, j-m, k-m}$ for any fixed $m$. We refer to entanglers for which the tensor $\mathbf{T}$ reduces to a single parameter as ultra-local.
Since we are interested in the theory in the vicinity of the critical point, we restrict our attention to entanglers that can generate a non-vanishing vacuum expectation value, as in Eq.~\eqref{veve}. This requires the inclusion of non-Gaussian entanglers that break the $\mathbb{Z}_2$ symmetry. In particular, we consider generators whose monomials have an odd total degree, ensuring that the induced field transformations map $\phi_n$ into even polynomials of the fields and can therefore generate a non-zero expectation value. Entanglers whose BCH expansion 
\begin{equation}\label{eq:transformations}
    e^{-i K[\phi, \pi]} \, \phi_n \, e^{i K[\phi, \pi]} = \phi_n - i \left[K, \phi_n \right] - \frac{1}{2!} \left[K, \left[K, \phi_n \right]\right] + \dots\, ,
\end{equation}
is infinite generally require computational techniques that extend beyond the operatorial framework employed in this work and may ultimately need to be analyzed on a case-by-case basis. Furthermore, in the presence of a nontrivial tensor structure 
$\mathbf{T}$, the explicit evaluation of the resulting infinite hierarchy of 
nested commutators rapidly becomes computationally intractable, and their 
analysis lies outside the scope of this paper. In what follows we focus on finite-BCH entanglers.

\subsection{Classes of finite-BCH polynomial entanglers}
We first consider entanglers whose exponent depends only on canonical conjugate momenta,
\begin{equation}
\label{Un}
U_C = \exp\Big(-i\sum_{m_1,\ldots,m_C} T_{m_1\cdots m_C}\,
\pi_{m_1}\cdots\pi_{m_C}\Big),
\end{equation}
where $T_{m_1\cdots m_C}$ is totally symmetric in its indices, with $C$ odd
so that the resulting field transformation can generate a non-vanishing
vacuum expectation value for $\phi$. The field and its conjugate momentum transform as
\begin{equation}
\phi_n \;\to\; \phi_n + C \sum_{m_1,\ldots}T_{n m_1\cdots m_{C-1}}\,
\pi_{m_1}\cdots\pi_{m_{C-1}}, \qquad \pi_n \to \pi_n,
\end{equation}
which follows directly from the BCH termination at first order discussed
above. Evaluating the resulting contribution to $\langle H\rangle$, however,
requires Wick-contracting up to $4\times(C-1)$ momentum operators on the
product state, so that the number of contraction terms grows
combinatorially with $C$; for this reason we restrict the explicit analysis
to $C=3$ and $C=5$, which already suffice to illustrate the structure of
the obstruction. The vacuum expectation value
$\langle\Phi\rangle$ is
\begin{equation}
\langle\Phi\rangle = \left\{
\begin{aligned}
&\frac{3}{2}\,\Delta\sqrt{2^r m_\phi}\sum_i T_{0ii}, & (C&=3),\\
&\frac{15}{4}\,\Delta^2\sqrt{2^r m_\phi}\sum_{m,q} T_{0mmqq}, & (C&=5),
\end{aligned}
\right.
\label{vevR}
\end{equation}
where we used translational invariance in the wavelet indices to set the first one to $0$, and used the product vev's
\begin{align}
    \langle \pi_n \pi_m\rangle = \frac{\Delta}{2}\delta_{nm}\,,\quad \langle \pi_n \phi_m\rangle = -\frac{i}{2}\delta_{nm}\,,\quad \langle \phi_n \phi_m\rangle = \frac{1}{2\Delta}\delta_{nm}\,.
\end{align}
In general, the tensor \(\mathbf{T}\) is characterized by a finite-range parameter \(\chi\), which controls its effective nonlocality range. More precisely, \(\chi\) is defined such that,
\begin{align}
    T_{m_1\dots m_C}=0\,,\quad {\rm whenever}\quad |m-m'|\geq \chi~,
\end{align}
where $m$ and $m'$ are an arbitrary pair of indices of \(\mathbf{T}\). The number of independent variational parameters of the non-Gaussian tensor  grows quadratically in $\chi$ due to the triangular structure of the tensor indices (e.g.
 \(\chi = 1,2,3,4,5\) correspond to
$
N_\text{var}=2,4,7,11,16,
$
respectively).

Similarly, a composition of the entangler $\exp(iK_2)$ with a Gaussian entangler \cite{wMERA}
\begin{equation}\label{eq:GS}
U_{GS} = e^{iK_{GS}}, \quad 
K_{GS} = -\frac{1}{2}\sum_{i,j} \zeta_{ij} 
\left(\phi_i \pi_j + \pi_i \phi_j\right),
\end{equation}
where $\zeta_{ij} = \zeta_{|i-j|}$ is a real symmetric kernel, leads to a BCH expansion for the transformation of $\phi_n$ and $\pi_n$ that is technically infinite but of the form
\begin{equation}
    \phi_n \to \sum_m (e^\zeta)_{nm} \phi_m + 3 \sum_{a,b} \tilde{T}_{nab} \pi_a \pi_b,
\end{equation}
where $\tilde{T}_{mbc}$ is an appropriate contraction of the original tensor 
$T_{mbc}$ with powers of $\zeta$. Although the BCH series is formally infinite, the transformed fields admit an exact closed-form expression obtained by resumming the nested commutators. Therefore, this class of entanglers is effectively equivalent to a finite-BCH entangler for the purposes of the present analysis. We emphasize that this finite BCH expansion is not the result of an  truncation of higher-order operator terms. Rather, for generators constructed solely from momentum (e.g. \(K_2\)) or field operators (e.g. \(K_3\)), the nested commutators terminate identically after a finite number of steps as a direct consequence of the canonical commutation relations. Although $K_3$ also generates a finite-BCH expansion, we disregard it because, since it acts trivially on $\phi_n$ ($\phi_n \to \phi_n$) and non-trivially on $\pi_n$, it generates a vev for the conjugate momentum and not for the field itself.

Entanglers obtained from $K_1$ generically lead to an infinite BCH expansion. This can be circumvented by engineering the tensor $T$ so that the BCH series terminates after a finite number of nested commutators, as also considered in \cite{IcMERA1,
InteractingcMERA3}.
  For instance, taking $K_1 = \sum_{i,j,k} T_{ijk} \phi_i \phi_j \pi_k + h.c.$, if $T$ is chosen such that the truncation occurs at first order, the generator induces the exact field transformations
\begin{equation}
\phi_n \to \phi_n +2 \sum_{k, q} T_{nkq} \phi_k \phi_q , \qquad \pi_n \to \pi_n - 2 \sum_{k,q} T_{kqn} (\pi_k \phi_q + \phi_q \pi_k),
\end{equation}
and the vev is of the form
\begin{equation}
    \langle \Phi \rangle = \frac{\sqrt{2^r m_\phi}}{2 \Delta} \sum_m T_{0mm} .
\end{equation}

For all the finite BCH entanglers considered above, the expectation value of the Hamiltonian in Eq.~\eqref{Hrtot} is a polynomial of degree at most four in the variational parameters,
\begin{equation}
\label{varR}
\langle H \rangle =
\frac{m_{\phi}^{2}}{2\Delta}
\left[ f_2(\mathbf{T}\Delta^{\frac{C}{2}}) + 2^{2r} h_2(\mathbf{T}\Delta^{\frac{C}{2}}) \right]
+ 2^r \frac{m_{\phi}^{3}}{\Delta^2} \,g
\left[ p_2(\mathbf{T}\Delta^{\frac{C}{2}}) + p_4(\mathbf{T}\Delta^{\frac{C}{2}}) \right].
\end{equation}
Here, $f_2$, $h_2$, and $p_2$ are quadratic monomials in the dimensionless quantity $\mathbf{T}\Delta^{\frac{C}{2}}$, whereas $p_4$ contains the quartic terms. 
We note that the mass dimension of $\mathbf{T}$ itself depends on the 
specific entangler under consideration, since it is fixed by the number 
of $\pi$ (or $\phi$) insertions in the corresponding generator. For $K_1$ entanglers with the tensor chosen so that the truncation occurs
beyond first order, Eq.~\eqref{varR} is not exact and instead contains
contributions of higher order in the variational parameters, up to order
$4k$, where $k$ is the truncation order. This does not affect the
argument below, which relies only on the fact that the vacuum expectation
value of the Hamiltonian is a polynomial of finite degree in the
variational parameters.
Once the entangler is chosen, the variational equations are solved for each choice of the coupling $g$ and mass $m$, determining the minimizing tensor $\mathbf{T}(g,m)$.
The resulting polynomial form of the energy functional \( \langle H \rangle \)  of Eq.\,\eqref{varR} is  exact within the class of ans\"atze we are focusing on, and is not a truncated approximation.

\section{An Algebraic Obstruction to Ising Criticality}

\label{sec:obstruction}

We first consider the simplest class of ultra-local entanglers.\,\footnote{This is realized only in the case of $e^{iK_2}$ without composition with the Gaussian entangler.}
The ultra-local $K_2$ entangler generates only Gaussian correlations and, on these grounds alone, cannot be expected to reproduce the critical behavior of the Ising universality class. Nevertheless, this simple setting provides a useful benchmark: it is analytically tractable and exhibits a second-order phase transition, allowing us to explicitly illustrate the emergence of mean-field critical behavior.
In the ultra-local case, the unique 
variational parameter $T$ can be traded for $\langle\Phi\rangle$, 
and Eq.~\eqref{varR} reduces to
\begin{equation}\label{ul}
\langle H \rangle =
C^2 \left(a_1 + g\, a_2^{(C)}\right) \langle \Phi \rangle^2
+ a_3 \frac{(4C-5)!!}{2^{2C-3}}\, C^4\, g\, \langle \Phi \rangle^4.
\end{equation}
For $C > 3$, the coefficient $a_2^{(C)}$ is negative and nontrivial 
solutions with $\langle\Phi\rangle \neq 0$ emerge. 
Eq.~\eqref{ul} is manifestly of Landau form, 
establishing mean-field critical behavior already at the ultra-local 
level. The case $C=3$ requires special treatment, since $a_2^{(3)} > 0$ 
forces the effective potential to have a unique minimum at 
$\langle\Phi\rangle = 0$, admitting only the trivial solution. 
This obstruction can be removed by augmenting the entangler with 
a shift operator $X_0 \sum_n \pi_n$, which acts as a constant 
displacement of the field and explicitly breaks the 
$\mathbb{Z}_2$ symmetry, allowing a nonzero vacuum expectation 
value to develop. The modified ansatz reads
\begin{equation}\label{U3}
\tilde{U}_3 = \exp\left[-i \left(T\sum_a \pi_a^3 + X_0\sum_n \pi_n \right)\right],
\end{equation}
where $X_0$ is an additional variational parameter. Under this 
ansatz, the field transforms as $\phi_n \to \phi_n + 3\,T\,\pi_n^2 + X_0$, so that $\langle\Phi\rangle = \sqrt{2^r m_\phi} (X_0 + 3T\Delta/2)$. Crucially, the addition 
of the shift operator does not alter the polynomial structure of 
the energy functional, so the Landau form of the effective 
potential and the mean-field obstruction are unchanged, as will become clear in the next subsection where we will consider an ansatz with an arbitrary number of variational parameters.

Naively, one would expect that systematically
increasing $\chi,$ thereby incorporating progressively
longer-ranged  non-Gaussian correlations, would
eventually drive the critical behavior away from mean field and
toward the exact Ising universality class. This expectation is based on the fact that finite-BCH entanglers generate genuine non-Gaussian correlations. For the entangler $U_3 U_{\rm GS}$, for
instance, the two-point function takes the schematic form
\begin{equation}
(G^{(2)}_{nm})_{\rm NGS} = (G^{(2)}_{nm})_{\rm GS} + \langle\phi_n\rangle \langle \phi_m \rangle +
\hat T_{nm}, \qquad \hat T_{nm} = \sum_{pq} T_{npq}T_{mpq},
\end{equation}
where $(G^{(2)}_{nm})_{\rm GS}$ denotes the Gaussian contribution. The
corresponding four-point function contains a connected piece built from all possible contractions of four tensors
$T_{nmk}$, that cannot be expressed solely in terms of products of
two-point correlators. This represents a genuine departure from Gaussianity. Nevertheless, as we show below, this is not sufficient to go beyond mean-field universality.

\subsection{The Landau Obstruction }
\label{4.1}
We now show that extending the variational ansatz beyond the ultra-local case does not alter its critical behavior. Increasing the nonlocality range $\chi$ introduces additional independent variational parameters in the tensor $\mathbf{T}$, but, generically, minimizing over them leaves the Landau behavior of the ultra-local case unchanged.
The first step is to fix one of the variational parameters and organize the remaining ones into a vector, which we denote by $\bm{Z}$. An obvious choice is to take $\langle \Phi \rangle$ as the fixed variational parameter. Since, as established in Sec. \ref{Epol}, $\langle H\rangle$ is a finite polynomial in the variational
parameters, the expectation value of the Hamiltonian takes the general form 
\begin{align}\label{eq:effective_potential}
\begin{aligned}
    \langle H \rangle 
&=
\langle \Phi \rangle\, \bm{a}\cdot\bm{Z}
+
\bm{Z}^T Q_2 \bm{Z}
+ g \;
\langle \Phi \rangle\, Q_3(\bm{Z},\bm{Z},\bm{Z}) + g \;
\langle \Phi \rangle^2\, \bm{Z}^T \tilde Q_2 \bm{Z}
\\
&\quad
 + g \;
\langle \Phi \rangle^3\, \bm{d}\cdot\bm{Z} +
g \; Q_4(\bm{Z}^4) + (b_1 + g \; b_2) \langle \Phi \rangle^2 + g \; b_4 \langle \Phi \rangle^4. 
\end{aligned}
\end{align}
The matrices $Q_2$ and $\tilde Q_2$ define quadratic forms in $\bm{Z}$, while $Q_3$ and $Q_4$ are rank-three and rank-four invariant 
tensors generating the corresponding interaction terms. The coefficients 
$b_1$ and $b_2$ are numerical constants. Both $\bm{a}$ and $Q_2$ 
consist of a $g$-independent term and a part linear in $g$. The specific coefficients and tensors depend on the chosen entangler, but their detailed form is not needed for our purposes.
We first consider the generic form of the variational energy
$\langle H\rangle$, in which the coefficients of the terms in
Eq.~\eqref{eq:effective_potential} are nonzero and $Q_2$ is non-singular
at the critical point. This condition is required for the
implicit-function-theorem argument used below. For the entanglers considered
in this work, we have verified numerically that
$\det Q_2\neq0$ at the critical point for $e^{iK_2}$ with $C=3$, for nonlocality ranges up to $\chi=20$. Although $\det Q_2$ may
vanish elsewhere in parameter space, this does not affect the argument,
which only requires $Q_2$ to be non-singular at the critical solution.
For completeness, we analyze separately the tuned case
$\det Q_2=0$ in Sec.~\ref{sec:degenerate}, where the
implicit-function-theorem argument breaks down, and show that this case
also cannot reproduce the Ising critical exponent.

To analyze the behavior of the $\bm{Z}$ parameters around the critical point, it is convenient to introduce
\begin{equation}\label{Feq}
    \bm{F}(\langle \Phi \rangle, \bm{Z}) \equiv \frac{\partial \langle H\rangle}{\partial \bm{Z}},
\end{equation}
with stationarity condition simply given by $\bm{F}(\langle \Phi \rangle, \bm{Z}) = 0$. By construction,
\begin{equation}
    \bm{F}(0,0) = 0,
\end{equation}
and the Jacobian with respect to the variational parameters satisfies
\begin{equation}
\left.
\frac{\partial \bm{F}}{\partial \bm{Z}}
\right|_{(0,0)}=
Q_2.
\end{equation}
Since $Q_2$ is non-singular,  the multivariate implicit function theorem guarantees the existence of a unique solution $\bm{Z}(\langle \phi \rangle)$ in a neighborhood of $\langle \Phi \rangle=0$, analytic in $\langle \Phi \rangle$. Sufficiently close to the critical point we can thus find a solution expanding $\bm{Z}$ in Taylor series for small $\langle \Phi \rangle$. The $\mathbb Z_2$ symmetry implies that if
$\bm{Z}(\langle\Phi\rangle)$ is a solution, then
$-\bm{Z}(\langle\Phi\rangle)$ is the solution corresponding to
$-\langle\Phi\rangle$. Since the local solution is unique, it follows that
\begin{equation}
\bm{Z}(-\langle\Phi\rangle)
=
-\bm{Z}(\langle\Phi\rangle),
\end{equation}
so that $\bm{Z}$ is an odd analytic function of
$\langle\Phi\rangle$. Therefore,
\begin{equation}\label{eq:solution_Ttilde}
\bm{Z}(\langle\Phi\rangle)
=
-Q_2^{-1}\bm a\,\langle\Phi\rangle
+\mathcal O(\langle\Phi\rangle^3).
\end{equation}
We now introduce the effective potential
\begin{equation}\label{eq:Veff}
V_\text{eff}(\langle \Phi \rangle)\equiv \min_{\langle \Phi \rangle=\text{const}}\langle H\rangle ,
\end{equation}
which, once we substitute Eq. \eqref{eq:solution_Ttilde} in the vacuum expectation value of the Hamiltonian, takes the Landau form
\begin{equation}\label{eq:Veff_Landau}
V_{\rm eff}(\langle \Phi \rangle)=
(a_2+g \; b_2)\langle \Phi \rangle^2
+
g \; a_4 \;\langle \Phi \rangle^4
+ \mathcal O(\langle \Phi \rangle^6),
\end{equation}
where $a_2$, $b_2$, and $a_4$ are appropriate combinations of the coefficients appearing in Eq.~\eqref{eq:effective_potential}. 
 Minimizing now $V_\text{eff}(\langle \Phi \rangle)$ close to the critical point we find $\langle \Phi \rangle \propto (g-g_c)^{1/2}$. The same result applies also in the (tuned) case in which $\bm{a} =0$: here we find $Z \propto \langle \Phi \rangle^3$, but the expansion of $V_\text{eff}(\langle \Phi \rangle)$ in powers of $\langle \Phi \rangle$ is the same as in Eq.\,\eqref{eq:Veff_Landau}, guaranteeing that $\beta = 1/2$. 
We conclude that the emergent Landau mean-field criticality is not a byproduct of a truncation approximation, but is instead a property of the exact variational ansatz, rooted in the analyticity of the effective potential in $\langle\Phi\rangle$ generated by finite-BCH polynomial operators. This shows that  
finite-BCH algebraic constructions remain confined to the same 
universality class regardless of the nonlocality parameter $\chi$, the 
number of variational parameters, or the resolution scale $r$. 
The argument applies to any finite-BCH entangler for which the vacuum
expectation value of the Hamiltonian is a finite polynomial in the
variational parameters. This includes the finite-BCH classes considered in
Sec.~\ref{Epol}, that is including $K_2$ and the engineered finite-BCH $K_1$
constructions. The argument
depends only on the lowest powers in the $\langle\Phi\rangle$ expansion and
is therefore independent of the total degree and coefficients of the
polynomial. Consequently, the same conclusion applies to $K_1$ entanglers
with truncation order higher than one.

The reader might wonder whether this argument continues to hold once the full
wMERA layering of Eq.~\eqref{vuotowMERA} is implemented. For the purpose of
establishing the obstruction to Ising criticality discussed above, however,
it is not necessary to solve the variational problem at each
wMERA layer. The obstruction follows from the polynomial structure
of the vacuum expectation value of the Hamiltonian induced by finite-BCH
entanglers, rather than from the values of the variational parameters obtained
through the layer-by-layer optimization. Thus, the fixed-layer analysis already establishes the obstruction, without requiring the full layering procedure of Eq.~\eqref{vuotowMERA}.

\subsubsection{Exceptions to mean field universality class}
\label{sec:degenerate}

We conclude with a comment on the degenerate case $\det Q_2 = 0$.
This case is not numerically realized anywhere in our analysis, and we do not expect it to be physically
relevant for the ansatz considered here. We nonetheless analyze it
explicitly because it represents the only way in which the
implicit-function-theorem argument of Sec.~\ref{4.1}
could in principle fail, and hence the only logical loophole through
which a finite-BCH entangler might evade mean-field behavior.

When one
or more eigenvalues of $Q_2$ vanish at the critical point, the implicit
function theorem argument breaks down. In the case of a single soft
direction $Z$ we can, however, use the Newton--Puiseux theorem to infer the
$\langle\Phi\rangle$ dependence of the variational parameter. The
stationarity condition along that direction is a polynomial relation
$F(x,y)=0$ between $x\equiv\langle\Phi\rangle$ and the soft component
$y\equiv Z$, with $\deg_y F\le3$ inherited from the cubic structure of
the derivative of Eq.\,\eqref{eq:effective_potential}. Once more, $F(0,0) = 0$ by construction. The Newton-Puiseux theorem states that a solution $y(x)$ of $F(x,y)=0$ near the origin always admits the fractional power series
\begin{equation}
y = \sum_n a_n\, x^{n/m}, \qquad m \le \deg_y F \le 3,
\end{equation}
with the exponent $n/m$ determined by the Newton polygon of $F$.\,\footnote{Given a polynomial $P(x,y) = \sum_{i,j} a_{ij} x^i x^j$, the corresponding Newton polygon is given by the convex region identified by the pairs $(i,j)$ appearing in $P(x,y)$. The leading exponent $n/m$ appearing in the Newton-Puiseux fractional series can be computed by simply taking the negative of the inverse slope of the segment of the Newton polygon that faces the origin.} Restricting Eq.\,\eqref{eq:effective_potential} to a single soft direction with corresponding $Q_2 = 0$, it is easy to show that the leading fractional exponent is given by $Z \propto \langle \Phi \rangle^{1/3}$. Substituting this back into Eq.\,\eqref{eq:effective_potential}, we obtain, close to the critical point,
\begin{equation}
    V_\text{eff}(\langle \Phi \rangle) = (a_{4/3} + g b_{4/3}) \langle \Phi \rangle^{4/3} + (a_2 + g b_2) \langle \Phi \rangle^2 + \mathcal{O}(\langle \Phi \rangle^{8/3}),
\end{equation}
which minimized with respect to $\langle \Phi \rangle$ gives $\beta = 3/2$. Although this is not mean-field, it is still far from the Ising universality class and, in fact, does not correspond to any known universality class. This observation, together with the fact that we do not encounter solutions satisfying $\det(Q_2)=0$ in any of the numerical variational calculations performed in this work, seems to suggest that this case should be considered more as a mathematical curiosity than a relevant physical situation. We observe that the condition $\det(Q_2) =0$ does not automatically imply $\beta \neq 1/2$. Consider, for instance, the tuned situation in which $a=0$ and $\det(Q_2)=0$ are true at the same time. Restricting once more to a unique soft direction, it is easy to see that the Newton polynomial collapses to a line of slope $-1$. This implies that $Z \propto \langle \Phi \rangle$ and we go back to an effective potential of the form of Eq.\,\eqref{eq:Veff_Landau}, with critical exponent $\beta = 1/2$. 

Finite-BCH polynomial
entanglers therefore fail to realize Ising criticality regardless of
whether the auxiliary sector is regular or singular. We have not exhaustively analyzed
configurations with multiple simultaneously vanishing eigenvalues of $Q_2$,
which would require a multivariate (Newton--Okounkov) generalization of
this argument.

\subsection{Numerical Illustration }\label{sec:numerical}
Since the analytical argument of Section\,\ref{4.1} relies only
on the polynomial structure of $\langle H \rangle$ and not on the
specific coefficients in Eq.\,\eqref{varR}, the critical behavior is
structurally identical across all finite-BCH entanglers by
construction. We now provide a numerical illustration of this result by restricting our analysis to the $C=3$ case, for which the entangler is given by:
\begin{equation}\label{eq:U3_total}
    \tilde{U}_3 = \exp\left[-i\left(\sum_{abc} T_{abc} \pi_a \pi_b \pi_c 
    + X_0 \sum_a \pi_a\right)\right].
\end{equation}

To demonstrate that the ansatz remains strictly within the mean-field universality class, we examine the critical exponent $\beta$ associated with the order parameter. Near the critical coupling $g_c$, the order parameter develops continuously according to
$ \langle \Phi \rangle \sim (g-g_c)^{\beta}, $ as $
     g \to g_c .$ 
Since universality classes are uniquely characterized by their critical exponents, the value of $\beta$ provides a direct and unambiguous probe of the universality class accessible to the ansatz.
 In particular, observing the mean-field value $\beta=1/2$ would indicate that the ansatz is confined to mean-field critical behavior, whereas a crossover toward the exact $(1+1)$-dimensional Ising value,
$ \beta = \frac{1}{8},$
would signal its ability to capture nontrivial critical correlations beyond the mean-field description.

Additionally, we show that increasing the number of variational parameters via $\chi$ systematically improves the variational energy while leaving the universality class unchanged. This occurs because
the polynomial structure of finite-BCH entanglers necessarily produces 
an effective potential of Landau form. Consequently, although the ansatz yields increasingly accurate variational energies, it remains unable to alter the underlying universality class.

We perform a numerical analysis of the variational equations associated with the entangler in Eq.~\eqref{eq:U3_total}, fixing $r=-1$ and considering system sizes up to $\chi \leq 20$, corresponding to up to 211 variational parameters. We also fix $\Delta = m_\phi$ to unity. In all cases, we observe a continuous (second-order) phase transition, in agreement with the  picture developed in the previous section.

\begin{figure}[t]
\centering
\includegraphics[scale=0.6]{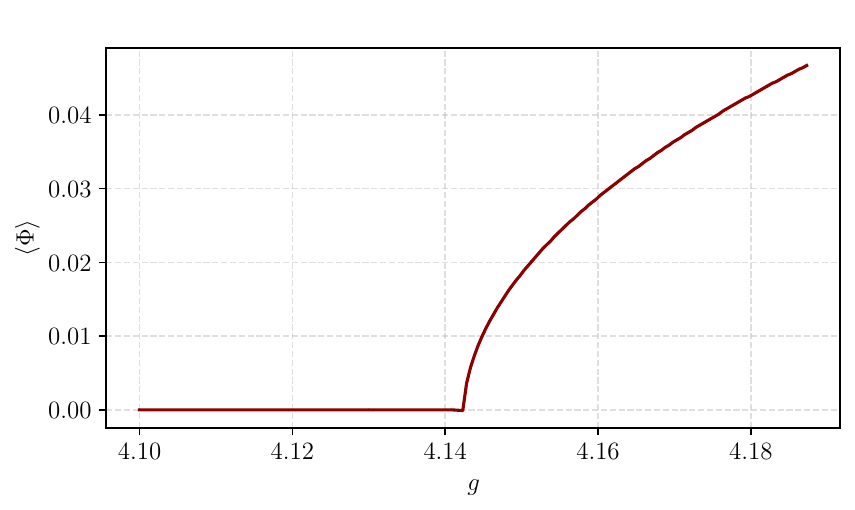}
\caption{
 Order parameter $\langle \phi \rangle$ as a function of $g$ for $\chi=20$ and $r=-1$.
}
\label{vev_vs_g}
\end{figure}

\begin{figure}[t]
\centering
\includegraphics[scale=0.6]{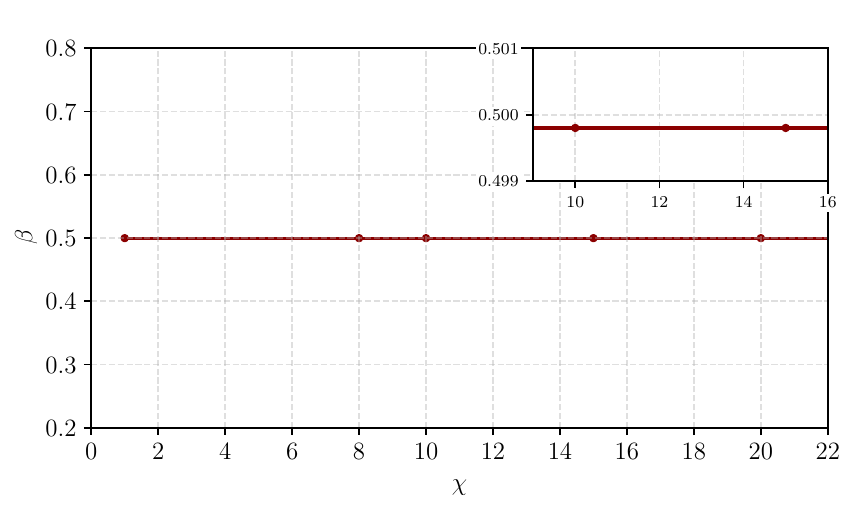}
\caption{ The extracted $\beta$ parameter as a function of $\chi$
 for $r=-1$. 
}
\label{beta_vs_chi}
\end{figure}
Figure~\ref{vev_vs_g} shows the vacuum expectation value of the field as a function of the coupling $g$ for $\chi=20$. To extract the critical exponent $\beta$ and quantitatively verify the expected Landau mean-field scaling, we perform a linear fit of $\log\langle\Phi\rangle$ as a function of $\log(g-g_c)$ along the non-trivial symmetry-breaking branch. For each value of $\chi$, the critical coupling $g_c$ is identified as the value of $g$ at which the Hessian of the variational energy $\langle H\rangle$, evaluated with respect to the variational parameters $\bm{Z}$ at $\langle\Phi\rangle=0$, first develops a vanishing eigenvalue, signaling the appearance of the symmetry-breaking instability. The fit is performed over the window
$
10^{-4} \leq g-g_c \leq 5\times10^{-3},
$
yielding critical exponents that are consistently centered around $1/2$ for all values of $\chi$ considered, as shown in Fig.\,\ref{beta_vs_chi}. In particular, for $\chi=2, 8,10,15,20$, the fitted exponent remains close to the mean-field value $\beta=1/2$, with no systematic dependence on $\chi$ and no indication of a drift toward the Ising value $\beta=1/8$. Since no formal uncertainty analysis was performed, the fitted exponent should be interpreted as a qualitative confirmation of the mean-field scaling rather than a precision determination of $\beta.$

\begin{figure}[t]
\centering
\includegraphics[scale=0.6]{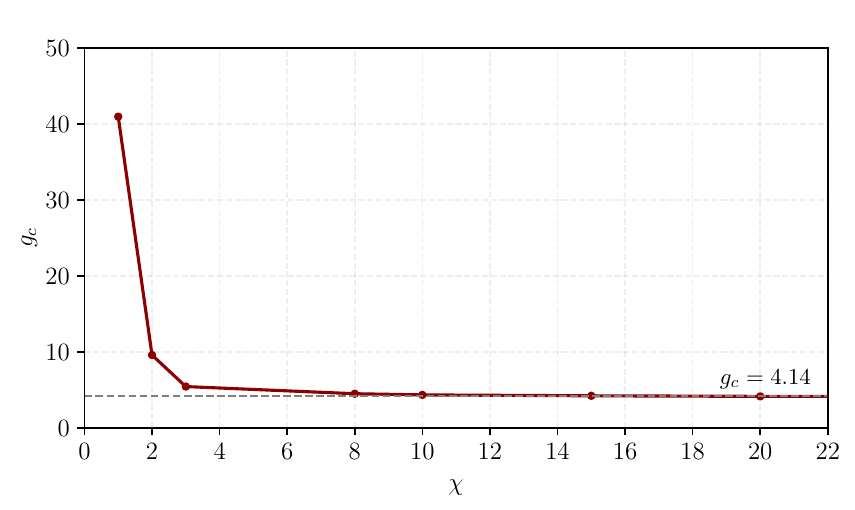}

\caption{
 Critical coupling $g_c$ as a function of $\chi$ for $r=-1$. }

\label{fig2}
\end{figure}

Figure~\ref{fig2}  shows the  critical coupling $g_c(\chi)$ for various values of $\chi$, which exhibits clear saturation for $\chi \gtrsim 10$, indicating convergence of the variational ansatz. 
We do not compare $g_c$ to literature values, since it is not a scheme-independent quantity, i.e.\ it depends on the details of the method used to compute it (for instance, the regularization scheme or the variational ansatz). This is consistent with the strong scheme-dependence of the critical coupling found even in continuum RG treatments of $\phi^4_2$.

\begin{figure}[t]
\centering

\includegraphics[scale=0.6]{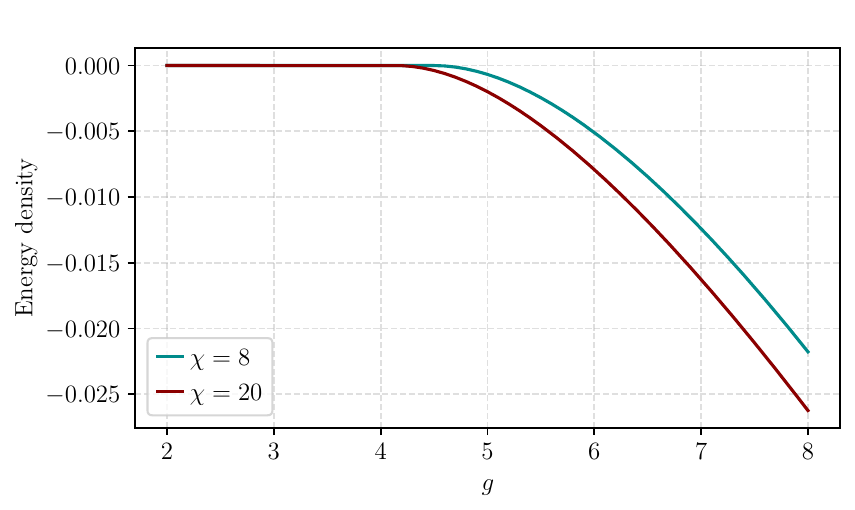}
\caption{
 Energy density as a function of $g$ for  $\chi=20$ (red) and $\chi=8$ blue.
}
\label{fig3}
\end{figure}
Figure~\ref{fig3}  shows the corresponding energy density as a function of the coupling. Increasing $\chi$  lowers the variational energy, indicating that the ansatz provides an increasingly accurate variational description.  Overall, increasing $\chi$ improves the variational energy and stabilizes non-universal quantities such as $g_c$, but does not modify the critical behavior, which remains consistent with mean-field scaling. 

As anticipated in Sec.~\ref{4.1}, we have also verified that the matrix
$Q_2$ introduced in Eq.~\eqref{eq:Veff} remains non-degenerate
throughout the parameter space explored numerically. Specifically, we find
that it is
bounded away from zero across the full grid of parameters explored in this work,
$-$ $\chi \leq 20$, $g \in (g_c - 1, g_c + 1)$, $r \in [-4,4]$ $-$ for entanglers with
$C=3$ and $C=5$. We have not extended this scan beyond $\chi = 20$, and we cannot
exclude the possibility that $\det Q_2 \to 0$ is approached only at larger $\chi$;
however, since  $g_c(\chi)$ remains stable in this regime
(Fig.~\ref{fig2}), we see no indication within the explored range that the
degenerate case is approached as $\chi$ increases further.

We also investigated the dependence of the results on the parameter $r$.  As expected from scale invariance in the vicinity of the critical point, varying $r$ does not affect the critical exponents, but merely shifts the value of the critical coupling. This behavior is confirmed numerically over the range $(-4 \leq r \leq 4)$. Since the corresponding results exhibit the same qualitative behavior and do not provide additional insight, we do not include the associated plots.

\section{Conclusions}
\label{conclu}
In this work, we investigated non-Gaussian entanglers within the 
wavelet multiscale entanglement renormalization ansatz (wMERA) 
applied to the interacting $\phi^4$ theory in $d=2$. We focused on the class of finite-BCH polynomial entanglers, which 
have been advocated as a tractable non-Gaussian framework~\cite{IcMERA1,InteractingcMERA3}, and 
carried out the first  analysis of their critical behavior.
Our analysis reveals an algebraic obstruction that limits the universality classes accessible to this entire family of entanglers. We showed that the effective potential 
generated by any finite-BCH polynomial entangler is necessarily of 
Landau form in the generic case, implying that the resulting critical 
behavior is governed by mean-field scaling, regardless of the 
nonlocality range or variational complexity of the ansatz. In 
non-generic, degenerate cases, the resulting critical exponent departs 
from mean-field ($\beta=3/2$) but still fails to reproduce the Ising 
value $\beta=1/8$, a possibility excluded numerically across all 
ans\"{a}tze considered. This conclusion is 
supported both by the analytical argument of 
Sec.~\ref{4.1} and by the numerical calculations 
of Sec.~\ref{sec:numerical}, where the critical exponent $\beta$ 
remains consistent with its mean-field value $\beta = 1/2$ with 
no drift toward the Ising value $\beta = 1/8$ as the variational 
complexity increases.
This obstruction is specific to the critical 
regime: the analytical argument relies on the behavior of the 
effective potential near $\langle\Phi\rangle = 0$ and does not 
affect the variational accuracy away from the phase transition. Finite-BCH entanglers could provide a potentially useful and computationally convenient framework in the non-critical regime, although this remains to be systematically investigated.

Our findings suggest that reproducing non-mean-field criticality,
and in particular the Ising universality class of $\phi^4$ in $d=2$,
requires going beyond variational ans\"atze whose effective potentials
have the Landau analytic structure established above, pointing toward
infinite-BCH entanglers or non-polynomial constructions as natural
directions. The simplest candidate for evading the polynomial
obstruction is an infinite-BCH $K_1$ entangler. However, even the
ultra-local example already exhibits significant difficulties. For
instance, consider
\begin{equation}
    U^{\text{(UL)}}_1 = \exp\left[-i\, T \sum_n \pi_n \phi_n^2\right],
\end{equation}
for which the field transformation resums to
\begin{equation}
    \phi_n \;\to\; \phi_n + \sum_{a=1}^{\infty} 2^a\, T^a\,
    \phi_n^{a+1}
    = \phi_n + 2T\,\frac{\phi_n^2}{1 - 2T\phi_n}.
\end{equation}
The vacuum expectation value takes the form
\begin{equation}\label{eq:phi_vev}
 \langle \Phi \rangle = \sqrt{\frac{2^{r+1} m_\phi}{\Delta}} \sum_a \frac{a!! \,2^{a/2} \,T^a}{\Delta^{a/2}} ,
\end{equation}
with the factorial growth implying divergence of the perturbative 
expansion. Even in this simplest ultra-local case, standard 
operatorial methods break down and non-perturbative resummation 
is required. Whether, once properly resummed, such entanglers can 
produce non-analytic contributions to the effective potential to 
thereby realize non-mean-field criticality, and in particular 
reproduce the Ising universality class of $\phi^4$, remains an 
important open question. Addressing this issue will require further investigation.


\acknowledgments
We thank Daniele Alves, Antoine Tilloy, Pietro Silvi and Simone Montangero for illuminating discussions. CF is especially grateful to Daniele Alves for numerous insightful and invaluable discussions on the wMERA framework.
EB and OC acknowledge the financial support of PRD 2026 funds at UNIMORE. EB also acknowledges partial support from the Italian INFN program on Theoretical Astroparticle Physics (TAsP). CF acknowledges partial support from the Italian INFN program on Precision Studies of Fundamental Interactions (SPIF).

\bibliographystyle{JHEP}
\bibliography{biblio}

\end{document}